# SHOCKS, SEYFERTS AND THE SNR CONNECTION: A *CHANDRA* OBSERVATION OF THE CIRCINUS GALAXY

B. Mingo [1], M. J. Hardcastle [1], J. H. Croston [2], D. A. Evans [3], P. Kharb [4], R. P. Kraft [3] and E. Lenc [5]
( Dated: Received ; accepted)
*Draft version September 3, 2012*

## ABSTRACT

We analyse new *Chandra* observations of the nearest ($D = 4$ Mpc) Seyfert 2 active galaxy, Circinus, and match them to pre-existing radio, infrared and optical data to study the kpc-scale emission. The proximity of Circinus allows us to observe in striking detail the structure of the radio lobes, revealing for the first time edge-brightened emission both in X-rays and radio. After considering various other possible scenarios, we show that this extended emission in Circinus is most likely caused by a jet-driven outflow, which is driving shells of strongly shocked gas into the halo of the host galaxy. In this context, we estimate Mach numbers $\mathcal{M} \sim 2.7$–$3.6$ and $\mathcal{M} \sim 2.8$–$5.3$ for the W and E shells respectively. We derive temperatures of $0.74^{+0.06}_{-0.05}$ keV and $0.8 - 1.8$ keV for the W and E shells, and an expansion velocity of $\sim 900$–$950$ km s$^{-1}$. We estimate that the total energy (thermal and kinetic) involved in creating both shells is $\sim 2 \times 10^{55}$ erg, and their age is $\sim 10^{6}$ years. Comparing these results with those we previously obtained for Centaurus A, NGC 3801 and Mrk 6, we show that these parameters scale approximately with the radio power of the parent AGN. The spatial coincidence between the X-ray and edge-brightened radio emission in Circinus resembles the morphology of some SNR shocks. This parallel has been expected for AGN, but has never been observed before. We investigate what underlying mechanisms both types of systems may have in common, arguing that, in Circinus, the edge-brightening in the shells may be accounted for by a $B$ field enhancement caused by shock compression, but do not preclude some local particle acceleration. These results can be extrapolated to other low-power systems, particularly those with late type hosts.

*Subject headings:* galaxies:active – galaxies:individual(Circinus) – galaxies:jets – shock waves – X-rays:galaxies –

## 1. INTRODUCTION

Recent *Chandra* observations of the environments of several radio galaxies (e.g., NGC4636, Jones et al. 2002; Hydra A, McNamara et al. 2000; Centaurus A, Kraft et al. 2003) have led to significant progress in understanding the AGN-driven gas outflows in these systems and the role they play in galaxy formation and evolution (as required by models such as those of Croton et al. 2006; Bower et al. 2006). We now know that, although the most powerful radio outflows, spanning hundreds of kpc, are associated with massive elliptical systems, smaller structures also connected to an active nucleus can be found in a variety of systems and environments, including spiral and disky galaxies (e.g. Gallimore et al. 2006; Kharb et al. 2006; Saikia & Jamrozy 2009).

The mechanism by which these structures are produced appears to be similar to the one we see in the most powerful sources, but on a smaller scale: the relativistic plasma ejected by the AGN interacts with the surrounding medium, pushing it out in kpc-scale radio-bright lobes, and inducing shocks that heat the medium to X-ray emitting temperatures (see e.g. Heinz et al. 1998; McNamara & Nulsen 2007; Shabala & Alexander 2009). The temperature structure of these bubbles has been analysed in a variety of systems, the most famous being Centaurus A, where the bow shock has been studied in detail after highly detailed images of the South-West lobe were obtained with *Chandra* (see e.g. Kraft et al. 2003; Croston et al. 2009).

The varied morphologies of the galaxies where this mechanism is found, and the fact that it is most likely episodic (Saikia & Jamrozy 2009; Kharb et al. 2006) make understanding the energetics involved in this process fundamental to estimating its impact on galaxy evolution. The thermodynamics of the expanding gas provide details on the energy output from the AGN, the radiative timescales involved in this energy being transferred to the ISM, and the way in which the energy input scales with the mass of the host galaxy, its morphology and the power output of the AGN, which also yields estimates of its lifetime. The injection rate, temperature and pressure of the gas determine star formation triggering and quenching in the central regions of the host galaxy.

Our previous results on systems such as Cen A (Croston et al. 2009), NGC 6764 (Croston et al. 2008b), NGC 3801 (Croston et al. 2007) and Markarian 6 (Mingo et al. 2011), show that gas outflows in these systems leave distinct signatures (thermal emission inside and in the rim of the lobes, synchrotron emission in very powerful shocks) whose physical properties indicate that the radio bubbles are overpressured with respect to their surroundings (as predicted by e.g. Capetti et al. 1999) and are, in several cases, driving shocks into them. For some of these galaxies (e.g. NGC 3801, Mrk 6) we have been able to calculate the (substantial) fraction of the AGN power that is being transferred to the ISM by this mechanism, setting the conditions for episodic AGN activity and star formation rates. Although the number of suitable sources for X-ray studies is limited, due to their small angular sizes,

[1] School of Physics, Astronomy & Mathematics, University of Hertfordshire, College Lane, Hatfield AL10 9AB, UK
[2] School of Physics and Astronomy, University of Southampton, Southampton SO17 1SJ, UK
[3] Harvard-Smithsonian Center for Astrophysics, 60 Garden Street, Cambridge, MA 02138, USA
[4] Department of Physics, Rochester Institute of Technology, Rochester, NY 14623, USA
[5] Sydney Institute for Astronomy, School of Physics, The University of Sydney, NSW 2006, Australia

| Obsid | Date | Exposure (ks) |
|-------|------|---------------|
| 00356 | 2000-03-14 | 25 |
| 12823 | 2010-12-17 | 160 |
| 12824 | 2010-12-24 | 40 |

for nearby galaxies this technique is particularly fruitful, since with long exposures we can even resolve different regions in the shocks driven by the lobes, allowing us to distinguish between thermal and non-thermal (synchrotron) emission.

In this paper we investigate the low-power AGN in the Circinus galaxy. Circinus is a nearby ($D$ = 4 Mpc) spiral galaxy, which exhibits a complex extended radio structure (Elmouttie et al. 1998a). While its nucleus and X-ray binary populations have been studied in some depth in the past (Smith et al. 2000; Smith & Wilson 2001; Sambruna et al. 2000, 2001a,b), the only preexisting X-ray observation where the extended emission was spatially resolved was too short to permit the study of the fainter emission. Our new, deep *Chandra* data allow us for the first time to study the X-ray counterparts of the larger-scale radio structures. Our main aim in the present paper is to test our new X-ray data, in conjunction with the existing radio observations, to test the different scenarios that might have created the extended emission we observe, and to derive the physical mechanisms and energetics involved.

Although Circinus is very close to us, the fact that it lies so close to our own Galaxy's plane has kept it from being observed in more detail in the past ($n_H = 5.56 \times 10^{21}$ cm$^{-2}$, Dickey & Lockman 1990). Even the distance to the galaxy is not very well constrained, with most authors relying on the estimation by Freeman et al. (1977) of $4.2^{+0.8}_{-0.8}$ Mpc, with slight variations. We have settled for what seems to be the most commonly accepted value, 4 Mpc, since this degree of uncertainty ($\sim$ 5 per cent variation in distance, $\sim$ 10 per cent in luminosity) should not have an impact on our conclusions, as we estimate it to be below the systematics.

## 2. DATA

We observed Circinus with *Chandra* in December 2010, for a total of 200 ks. We also used a pre-existing observation from 2000, originally analysed by Smith & Wilson (2001) (see Table 1), adding up a total of 225 ks. We reduced our data using the standard CIAO 4.2 pipeline, applying the latest calibration files. The observations show no trace of background flares, so that we could use the full exposures for all of them.

We generated event files for each *Chandra* observation, and merged them into a single image with the CIAO tool *merge_all*. We identified the point sources with *wavdetect* and cross-checked them manually for false or omitted identifications. We then removed the point sources and the readout streak (which is very prominent in the longest observation, running roughly in the same orientation as the galaxy's disk) from the merged image, to keep them from interfering in the analysis and to achieve the best possible characterization of the extended emission. Finally, we filled in the resulting gaps with the CIAO tool *dmfilth*. The resulting image can be seen in Fig. 1.

We used radio maps from archival *ATCA* (Australia Telescope Compact Array) observations at 21, 13, and 6 cm. The analysis of the original data is described by Elmouttie et al.

(1998a), who also discuss the radio properties of the source in detail.

In addition to the data of Elmouttie et al. (1998a), we also obtained wide-band (2 GHz) *ATCA* data collected during CABB (Compact Array Broadband Backend, Wilson et al. 2011) commissioning and as part of an ATNF Summer Student observation of Circinus. The CABB commissioning data were taken on 02 April 2009 ($\sim$8 hr, 6 cm only) using a compact 352 m configuration and 13 April 2009 ($\sim$6 hr, 3 cm and 6 cm) using a compact 168 m configuration. The ATNF Summer Student observations were made on 16 January 2010 ($\sim$9 hr, 3 cm and 6 cm) using a 6 km configuration.

The *ATCA* CABB data were calibrated in the MIRIAD package (Sault et al. 1995) and flagged using the new MIRIAD task mirflag. Model fitting of the source was performed using uv-components in DIFMAP (Shepherd et al. 1994) with an additional parameter used to model the spectral index of each component (thus accounting for spectral variation of the source across the 2 GHz wide band). Both phase and amplitude self-calibration were performed iteratively to improve the calibration and subsequent models. The final model achieves a 1 sigma sensitivity of 50 $\mu$Jy/beam.

We also compared our data to the results on the ionized gas of the galaxy (Elmouttie et al. 1998b), the HI emission (Jones et al. 1999; Curran et al. 2008) and the optical images of the centre of the galaxy obtained with the *HST* (Wilson et al. 2000).

We used XSPEC 12.5 to fit models to the X-ray spectra. All the model parameters thus estimated are quoted with 90 per cent confidence uncertainties.

## 3. ANALYSIS

### 3.1. *Imaging*

The reduced and merged X-ray image we obtained is presented in Fig. 1. Overlaid over the X-ray image are the 13 cm radio contours and the regions we used to extract the surface brightness profiles shown in Figs. 3 and 4. This image shows several extended structures, with different orientations (see also Curran et al. 2008). Our proposed geometry is detailed in Fig. 2, which was drawn over the combined X-ray/radio/optical image, to reflect the scale, reach and orientation of each component as accurately as possible. Our Figure is similar to the picture proposed by Elmouttie et al. (1998a), but there are some differences, particularly in the interpretation of the different structures. The AGN and circumnuclear star-forming ring are at the centre of the galaxy, the ring facing towards us. The galaxy's disk extends in the NE-SW direction, with an orientation of $\sim 60°$ relative to the plane of the sky (Jones et al. 1999; Curran et al. 2008) and shows some X-ray emission, though it is fainter than the other structures. The NW lobe is unobscured, on the near side of the disk, and the SE lobe is partly obscured, showing a clear dip in X-ray emission in the areas covered by the disk. Both lobes appear to be edge-brightened, though the NW lobe is much brighter and somewhat smaller. The orientation of the lobes is roughly perpendicular to that of the disk. There is a dip in X-ray luminosity on both lobes behind the bright edges, and a rise towards the centre of both, suggesting that there is further structure within the lobes. We extracted surface brightness profiles for both lobes, to quantitatively verify the correspondence between the X-ray and radio structures, as well as to confirm the edge-brightening in both cases. The results are illustrated in Figs. 3 and 4. Near the AGN, coinciding with the base of the NW lobe, quite strong X-ray emission coincides with a region of highly ionized gas that is believed to be an ionization cone (Marconi et al. 1994; Elmouttie et al. 1998b; Wilson et al. 2000; Smith & Wilson 2001). The X-ray emission from the galaxy's halo is very faint and not directly discernible from the images shown.

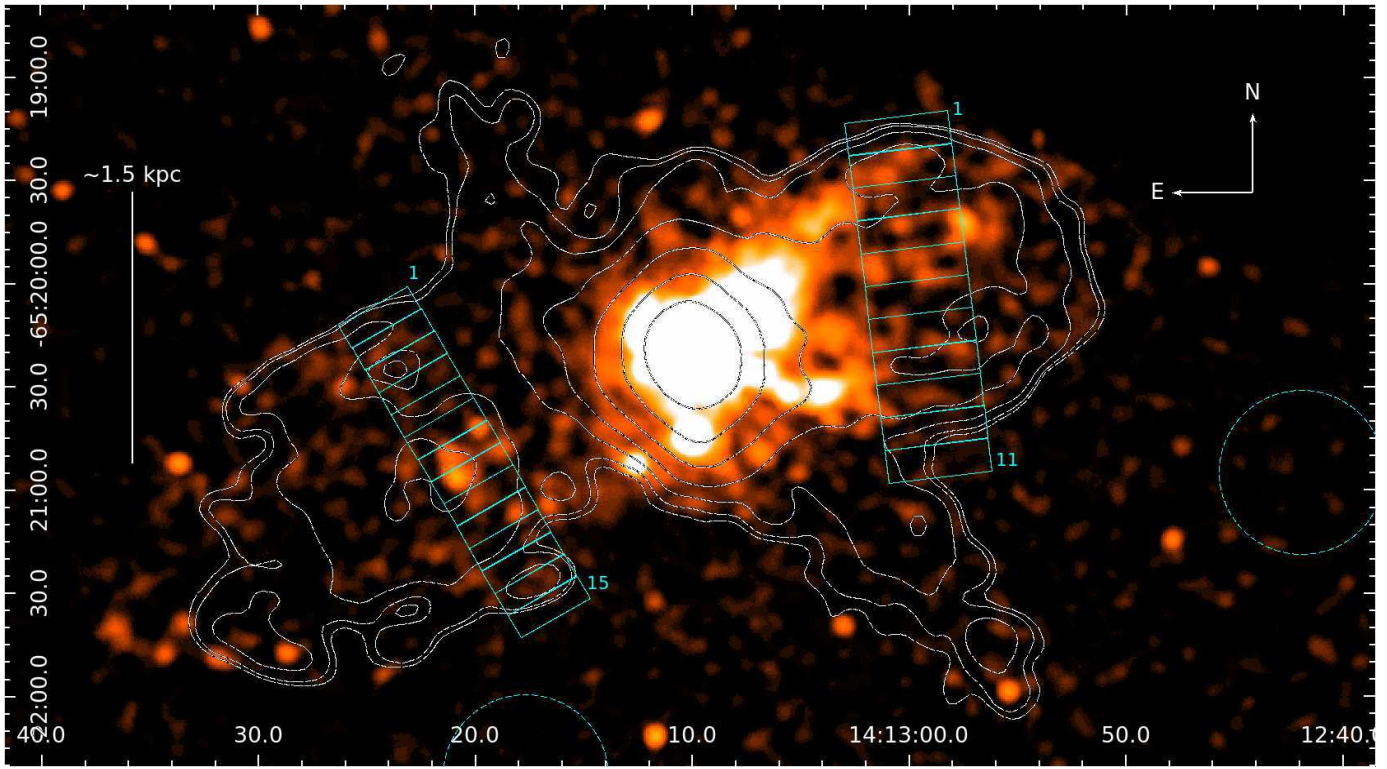

**Figure 1.** Merged (new+archival ACIS-S observations) 0.4-5 keV *Chandra* emission with overlaid *ATCA* 13 cm contours and $\sigma = 8$ *Chandra* pixels (4 arcsec) Gaussian smoothing, displaying the extent and morphology of the extended emission. Any point sources except the nucleus and the SNR have been masked out. The X-ray emission coincides with the edges of the radio lobes (NW and SE structures), although there is some diffuse X-ray emission coinciding with the galaxy's disk (extending NE-SW). This image also shows the regions we used to extract surface brightness profiles for both radio lobes (the background was chosen from the circular, dashed regions). The numbers indicate the first and last regions in each strip. Scale: 19.4 pc arcsec$^{-1}$. The resolution of the 13 cm radio map is $11.8 \times 11.0$ arcsec.

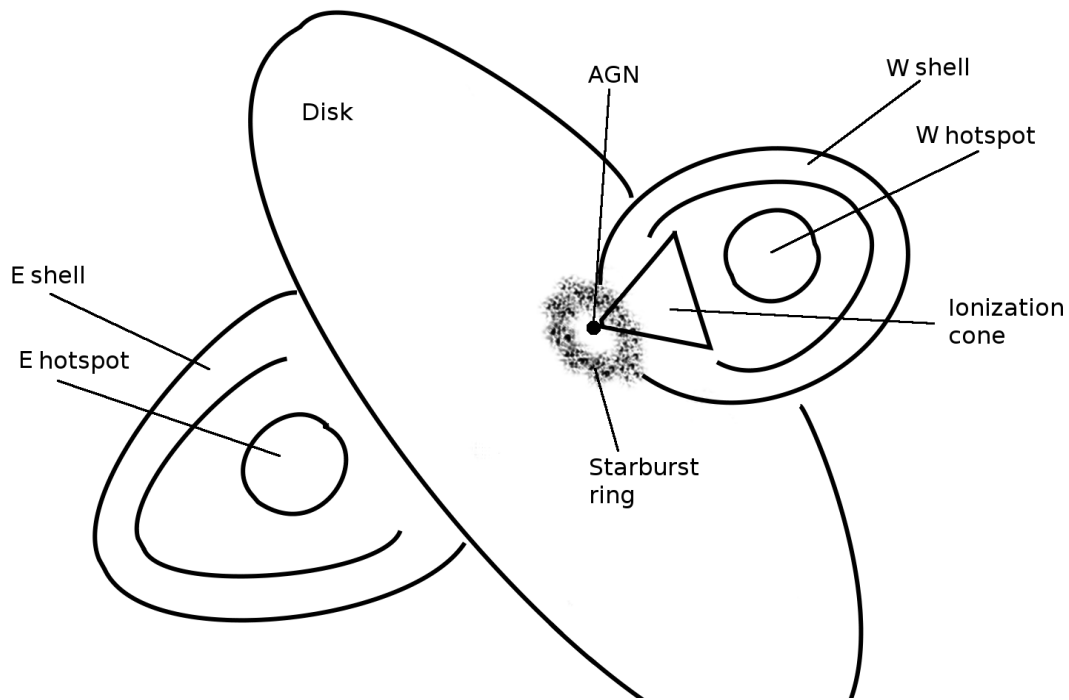

**Figure 2.** Scheme of the different components we identify in the X-ray and radio images.

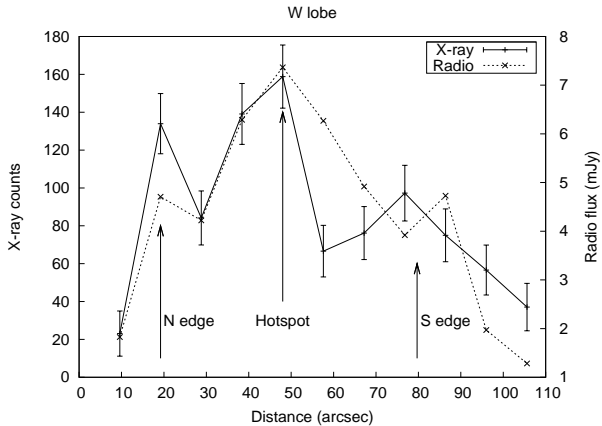

**Figure 3.** Surface brightness profiles for the W lobe of Circinus, as extracted from the regions in Fig. 1. The regions are $30 \times 9.6$ arcsec in size, and have an inclination angle of 6.9°. The distance is measured along the regions in the N-S direction, starting on the northernmost edge of the first region. The arrows indicate the N and S edges of the shell, and the hotspot, as illustrated in Fig. 2. Scale: 19.4 pc arcsec⁻¹

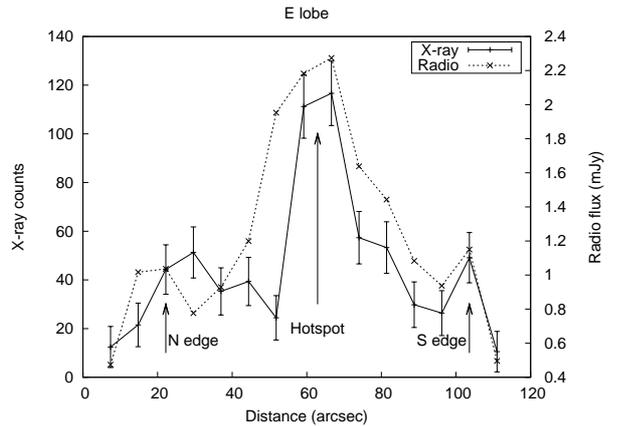

**Figure 4.** Surface brightness profiles for the E lobe of Circinus, as extracted from the regions in Fig. 1. The regions are $23 \times 7.4$ arcsec in size, and have an inclination angle of 30°. The distance is measured along the regions in the N-S direction, starting on the northernmost edge of the first region. The arrows indicate the N and S edges of the shell, and the hotspot, as illustrated in Fig. 2. Scale: 19.4 pc arcsec⁻¹

The radio contours show the overlap between radio and X-ray emission (for a detailed analysis of the radio maps see Elmouttie et al. 1998a). There is faint radio emission in the NE-SW direction, coincident with the disk of the galaxy, while the radio lobes extend perpendicularly to it. The pattern of radio emission in the lobes matches the X-ray structures perfectly: the edge-brightened emission in the lobes, the dip in luminosity behind the edges and the rise towards the centre of the lobes are spatially coincident across both wavelengths, except in a few points. We tested different orientations and distances when extracting the radial profiles in Figs. 3 and 4. The coincidence between the radio and X-ray edge peaks persists across most of both lobes, though it is clearer in some areas. We chose regions that covered all the appropriate structures: shells, lobe interior and hotspots, while trying to avoid the main emission from the ionization cone, which has no radio counterpart, and the edges of the CCD.

The differences in the morphology of the emission between the two lobes can only be partly explained by orientation effects. The E lobe is larger, and does show X-ray emission outside the confines of the radio lobe edges in the regions farthest from the AGN. The W lobe is smaller, more spherical, and its emission seems to be confined within what is seen in the radio maps. However, the edge of the CCD is just outside the edge of the W lobe in our long observations, and it is possible that we may be missing some of the outermost edge in this direction, in which case both lobes could exhibit a tip of X-ray emission outside the radio lobe. It is possible that the outflow may have encountered very different environments on its way to its current position, which could explain the asymmetry between the W and E lobes.

The W structure also shows the base of the ionization cone detected by Marconi et al. (1994), while no counterpart is visible in the E, most likely due to its being obscured by the galaxy's disk (see Figs. 5 and 2). The *Chandra* images show three distinct tails, propagating from the AGN in the W direction. The two brighter tails to the North have [OIII] and Hα counterparts (see the right panel of Fig. 5 and Smith & Wilson 2001; Wilson et al. 2000; Elmouttie et al. 1998b) while the southernmost edge only shows up in Hα + [NII]. This southernmost edge could be obscured, as suggested by Elmouttie et al., or could be produced by a different mechanism. While the top tail is clearly one of the edges of the ionization cone, it is unclear whether the other edge is given by the central or the southernmost tail (Elmouttie et al. 1998b).

### 3.2. Spectroscopy

As we discussed in Section 1, the nucleus of Circinus and its immediate surroundings have been studied in detail in the past. The objective of our analysis is to study the larger scale X-ray emission associated with the radio lobes of Circinus.

We extracted X-ray spectra of the radio lobes from regions excluding the AGN and circumnuclear emission, as well as the (presumably) photoionized plume in the W lobe, to avoid contamination. Our W region also avoids most of the area covered by the part of the disk that lies behind the lobe, thus minimising any possible contamination from it in the resulting spectrum. The spectra were binned to 15 counts per bin to allow Gaussian statistics to be used. We also binned the spectra to 20 counts per bin to check the robustness of the fits.

We used regions that cover most of the radio lobes, excluding the circumnuclear emission and any point sources, after verifying that the spectrum of the gas inside the lobes does not differ from those of the gas in the shells, presumably due to the low surface brightness emission from the shells contributing in these regions. We worked with a number of different source and background regions to minimize contamination. The spectra of the W lobe are fairly consistent regardless of our choice of source and background, but the E lobe was problematic in this respect: we found that the overall shape of the spectrum was independent of our choice of regions, but not entirely consistent with any single or double component model. As discussed below, this anomalous shape may be caused by a higher contribution of star-related emission from the galaxy's disk. The statistics on the E lobe are also poorer than those of the W counterpart, most likely due to the higher obscuration, making it difficult to mask out bright regions and still obtain a good fit.

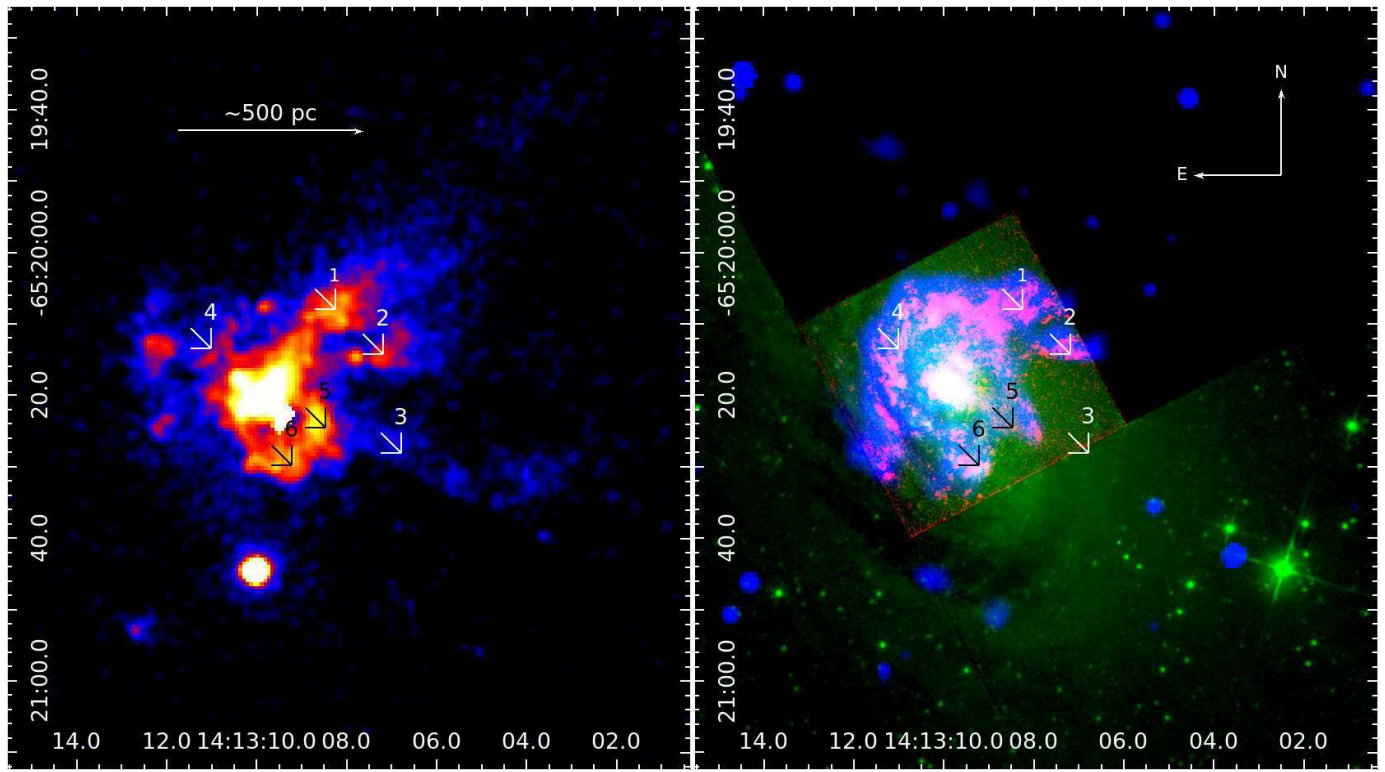

**Figure 5.** Left: The central regions of Circinus as seen in our merged, masked Chandra image, with $\sigma = 2$ pixel (1 arcsec) Gaussian smoothing. The image shows the AGN, the circumnuclear star-forming ring and the ionization cone at the base of the W radio lobe. Right: composite image showing the *HST* data on [OIII] emission (red), and F814W filter (broad I) continuum (green) from Wilson et al. (2000), and the H$_\alpha$ emission (blue) from Elmouttie et al. (1998b), kindly provided by Bi-Qing For, illustrating the correspondence between the structures we observe in X-rays and those visible in the optical. The arrows point to matching structures: the N tail of the ionization cone (1,2), the S tail (3), which is obscured in the optical images, and parts of the circumnuclear ring (4, 5, 6). See Fig. 2 for more details on the geometry around these structures. Scale: 19.4 pc arcsec$^{-1}$

We fitted to our spectra a model consisting of local Galactic absorption (wabs) and emission from a collisionally ionized gas (apec) (see Figs.6 and 7). We approached the analysis by fitting the model both to the individual spectra together and to the coadded spectrum (generated with the CIAO tool *combine_spectra*). Only the former approach is shown in the Figures, but the results are consistent for both methods. Given that the abundance is not well constrained (especially for the E lobe, where we have problems with both multiphase gas and covering absorbers), and if left free during the fitting the values tend to become implausibly low, we assumed a fixed abundance of 0.15, compatible with the lower limit set by the abundances in the IGM (see e.g. Danforth et al. 2006). Abundances higher than 0.2 do not affect the estimated values for the temperature, but the fit statistics are much worse. See the last paragraph of Section 4.2 for a discussion on the implications of a different value of $Z$.

We found that the emission in the W lobe is very well fitted with this single temperature model, although there are some residuals, especially at high energies, most likely caused by contamination from the AGN PSF. We obtained a best-fitting temperature of $0.74^{+0.06}_{-0.05}$ keV, and a reduced $\chi^2$ of 1.09. See Fig. 6 and Table 2 for details.

The E lobe, however, proved to be more problematic. This area is partially covered by the galaxy's disk, and the spectrum (see Fig. 7) is not successfully modelled by a single temperature model (reduced $\chi^2 \sim 1.5$). Adding a powerlaw or a second thermal component does improve the fit, albeit slightly (reduced $\chi^2 \sim 1.2$) and most of the residuals persist. Adding an extra absorbing column does not improve the fit.

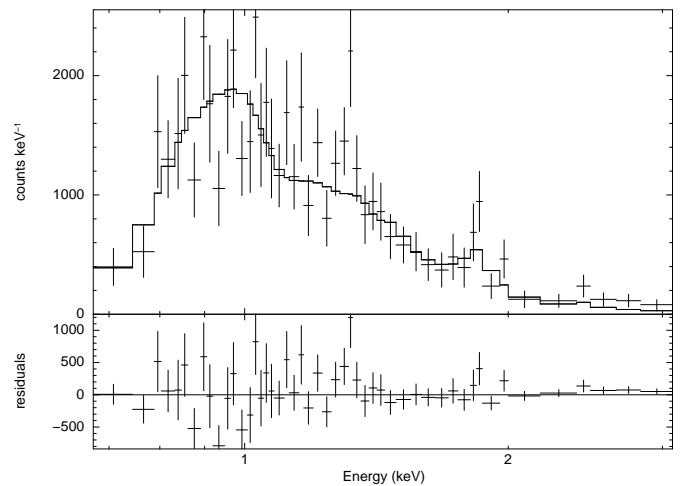

**Figure 6.** Spectral fit to the combined spectrum of the W lobe of Circinus. Model: wabs*apec, $kT = 0.74^{+0.06}_{-0.05}$ keV.

The poor fit could be caused by emission from an unresolved X-ray binary population and massive stars in the disk, or unavoidable contamination from the nearby, strong AGN.

XSPEC determined a best fitting temperature for the E lobe of $kT = 1.6^{+0.6}_{-0.2}$ keV, but it is not well constrained. A detailed analysis shows local minima when the model assumes $kT = 0.8$ and $kT = 1.8$ keV. Thus we used these values as a lower and upper limit, respectively, for the shell calculations displayed in Table 2.

The disk over the E lobe is problematic not only because it may be introducing some contamination, it might also be masking an E counterpart to the W ionization cone, which could also be contributing to our spectra. Given the different morphologies of the two lobes, it is also possible that the temperature structure may be different for the E structure. We will discuss the nature of the emission in the lobes and its structure in the following Sections.

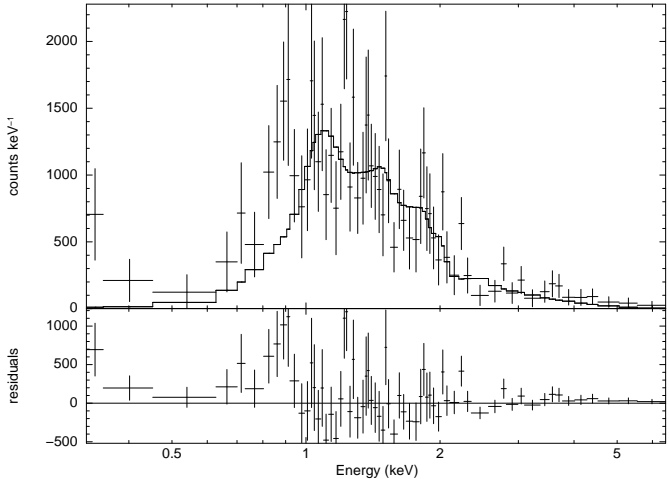

**Figure 7.** Spectral fit to the combined spectrum of the E lobe of Circinus. Model: wabs*apec, $kT = 1.6^{+0.6}_{-0.2}$ keV.

We also carried out a fit over the regions corresponding to the disk (NE-SW), and found that a single temperature model works remarkably well, with $kT = 0.69^{+0.19}_{-0.17}$ keV. This temperature is very similar to the one we found for the W shell, but not so well constrained (mainly due to a higher fraction of contamination from the AGN).

To test whether there are any changes in the nature of the emission in the brighter areas within the lobes, we extracted spectra from these regions. A fit to a single powerlaw model gave very poor results, since the thermal emission from the shells is still present. We then fitted to the spectra a model equal to the one we used for each shell plus either a powerlaw or a second thermal component. The addition of a second component improves the fit; however, the poor statistics (the brightest individual spectrum has just ∼100 counts) do not allow us to distinguish between a thermal and a non-thermal model in this case. Although the fit is slightly better with the non-thermal model, the difference is not statistically significant. We will discuss the nature of this emission in Section 4.3.

## 4. DISCUSSION

In this Section we discuss the possible scenarios that may be causing the emission we observe. In Section 4.1 we discuss the possibility that the large-scale emission is caused by photoionization from the AGN. In Section 4.2 we argue that what we are observing is, instead, a shock. As possible causes behind this shock, we contemplate the possibility of a starburst-driven superbubble (Section 4.3) or a radio jet-ISM interaction (Section 4.4).

### 4.1. Photoionization

Given that the edges of the ionization cone coincide with the base of the W shell almost exactly (see Fig. 5 and the discussion at the end of Section 3.1), it is possible that some of the line emission studied by Marconi et al. (1994) may be caused by collisional ionization rather than photoionization. However, without [O III], [N II] and H$\alpha$ maps with a large field of view and good resolution to measure the intensities and ratios, it is hard to tell. Given the intensity of the AGN and the light from the starburst ring, photoionization must indeed play a role in the emission we see in this region. However, it is extremely unlikely that photoionization from the central AGN could explain the edge-brightened, large-scale emission we are observing further out. We calculated the extent of photoionization caused by the AGN following the methods of Wang et al. (2009) and the models from Kallman & McCray (1982). Given that:

$$\xi = \frac{L}{nR^2} \qquad (1)$$

where $\xi$ is the ionizing parameter, which we assume to be between 10 and 100 (reasonable values for AGN outflows, as demonstrated by Ogle et al. 2003, see also Wang et al. 2009). $n \sim n_e$ is the particle density (which we assume to be between $10^{-2}$ and $10^{-1}$ cm$^{-3}$), $R$ is the distance (in cm) and $L$ is the unabsorbed X-ray luminosity of the AGN required to produce the observed ionization factor (we assumed $L = 1.6 \times 10^{42}$ erg s$^{-1}$, after de Rosa et al. 2012). In the case of Circinus, this implies that the AGN would be able to weakly ionize ($\xi \sim 10$) gas up to $\sim 700$ pc away, but strong ionization ($\xi \sim 100$) would only be produced within the inner $\sim 200$ pc. Given the relative low luminosity of the AGN (Yang et al. 2009) and the dense environment surrounding it, this is consistent with what is observed for the ionization cone, but shows that the edge-brightened emission in the shells cannot be accounted for by photoionization. Given that the X-ray emission at the edges of the radio lobes extends up to 2-3 kpc in projection (for the W and E shell respectively), even the possibility that the AGN was more powerful $10^3$–$10^4$ years ago makes photoionization unrealistic to explain the level of emission we are detecting at these distances, requiring densities an order of magnitude smaller than those we measure in these regions to achieve weak photoionization.

While there are no high resolution, large scale [O III] or H$\alpha$ maps that cover all of the ionization cone, the existing images (Elmouttie et al. 1998b; Wilson et al. 2000) do show other regions of the galaxy that coincide with part of the lobes, and there seems to be no enhancement in emission in these regions, matching the radio and X-ray data, while this enhancement is observed in the ionization cone. From a morphological perspective, since the base of the W shell coincides with the ionization cone (Marconi et al. 1994), and given the results above, we can infer that there is a transition between predominantly photoionized and shock-ionized gas at these distances, which could be further investigated with large scale [O III] and H$\alpha$ maps and more X-ray observations.

For consistency, we extracted a spectrum of the two northernmost tails of the ionization cone (see left panel of Fig. 5) in our longer exposure, covering distances of 150–500 pc from the AGN. We fitted the resulting spectrum to a *photemis* model (part of WARMABS, an analytical XSPEC model implemented from XSTAR) and foreground Galactic absorption. We assumed a gas density of $10^4$ cm$^{-3}$ and an ionizing powerlaw with $\Gamma \sim 1.6$ (de Rosa et al. 2012). We found that photoionized emission alone does not fully account for the spectral shape, yielding a reduced $\chi^2 \sim 2.5$. This may be due to the less than ideal background subtraction (a local background region that can compensate for the AGN contamination, without including the readout streak or emission from the star-forming ring will still include thermal emission from the galaxy's disk) or to a more complex spectral structure de-rived from the large volume covered by our region (a requirement to obtain the necessary statistics for the fit). By testing the addition of different continua to improve the fit, we observed that the ionization parameter remained stable, at values of $\log(\xi) \sim 1.7$–2.0, which supports our assumptions in the previous paragraph. We also found very low abundance constraints ($\sim 10^{-2}$) on O, Mg and Si, and higher values ($\sim 0.1$–0.3) for Fe and Ne. We were unable to constrain the abundances for the other elements, but increasing them over our initial assumption (0.15$\odot$) increases the $\chi^2$.

We also studied the likelihood of detecting an identical structure in the E lobe, behind the disk of the galaxy. We estimated the countrate we would expect in our image from this structure with the PIMMS tool, assuming it to have the same unabsorbed flux as our W ionization cone spectrum ($F_{0.3-10keV,unabs} = 1.5 \times 10^{-12}$ erg cm$^{-2}$ s$^{-1}$) and a foreground absorbing column from the disk comparable to that of the Cen A dust lane ($N_H \sim 1.5 \times 10^{22}$ cm$^{-2}$). We found that we would only obtain $\sim$300 counts from this structure in our image, below the emission level from the disk of the galaxy itself, and thus we would not be able to detect it. Although the obscuration from the disk might be lower, it is also possible that the E ionization cone, if present, may be fainter than its W counterpart, if the environment through which the radiation must pass is denser. The possibility of an ionization cone in the E lobe, therefore, cannot be discarded.

### 4.2. A shock model

The match between the X-ray and radio emission is striking in Circinus (see Figs. 3 and 4). Unlike in other, more distant systems, where we lack the spatial resolution to resolve these structures in such detail, the proximity of Circinus allows us to observe that the emission from the shells coincides extremely well in radio and X-rays. In all the radio frequencies there is a dip in luminosity behind the shell, and a rise in emission towards the centre of the radio lobes. Given the edge-brightened nature of the emission, the abrupt jump in luminosity and the spectral properties we derived for the gas, we consider that the most likely scenario is that we are observing shells of shocked gas around the radio lobes of Circinus. We will discuss below whether this shock may be starburst or jet-driven. The dip and rise in luminosity behind the shells suggests further complexity in their structure, since we would expect the lobes to be uniformly filled with gas. We will discuss the nature of this emission for both the starburst and jet scenarios.

The fact that the X-ray emission does not appear to be outside the radio lobes (except at the tip of the E lobe, as discussed in Section 3.1) suggests that the same shock that is causing the X-ray emission in the shells (which, as detailed in Section 3.2, is thermal in origin) is also giving rise to enhanced synchrotron emission at radio frequencies. The apparent thickness of the shells, measured from the X-ray images, is $\sim$ 300 pc, and appears constant across the entire structure.

To test the shock hypothesis we assumed two ellipsoidal shells (the geometry that most closely follows the shells as depicted in Fig. 2, with the smallest departure from spherical symmetry) with the two minor axes equal. The two-dimensional regions we used in our images have eccentricities of 0.74 and 0.52 for the E and W shells respectively. The E structure is larger, with $R_{int,maj} = 1.0$ kpc, $R_{ext,maj} = 1.4$ kpc, $R_{int,min} = 0.7$ kpc, $R_{ext,min} = 0.9$ kpc. The W shell has $R_{int,maj} = 0.8$ kpc, $R_{ext,maj} = 1.1$ kpc, $R_{int,min} = 0.7$ kpc, $R_{ext,min} = 0.9$ kpc. These dimensions correspond to volumes of $7.8 \times 10^{64}$ cm$^3$ and $6.1 \times 10^{64}$ cm$^3$ for the E and W shells respectively.

We derived the electron densities for both shells from the normalization of the thermal model (see Table 2), assuming a uniform filling factor. From these densities it is also possible to calculate the total mass of gas in the shells, the pressure, the total thermal energy, and the work available from the gas. We found the densities to be slightly higher than those we found for Mrk 6, but the total energy involved is lower by roughly two orders of magnitude, which is to be expected given the much lower radio power observed in Circinus.

In order to test our hypothesis of a shock scenario we attempted to extract spectra of the gas immediately outside the lobes, but were unable to fit a model to the data, due to the poor statistics. We then generated radial profiles for the longest observation from concentric annuli ($r_{min} = 132$ arcsec, $r_{max} = 305$ arcsec, $\delta r = 8.7$ arcsec), excluding the emission from the AGN, the lobes and the galactic disk, and a suitable background. The surface brightness of the gas in the halo is extremely low, and it barely shows up in radial profiles (see Fig. 8). The drop in luminosity at R$\sim$ 440 pixels ($\sim 4.2$ kpc) indicates the possible extent of the halo, but the errors are too large to consider this anything more than a qualitative indication.

To better constrain the surface brightness we estimated the count statistics for a circular region centered in the AGN in the longest observation, excluding the regions coincident with the lobes, the disk of the galaxy, any point sources, the readout streak and any CCD gaps or adjacent chips in the process. We used two different radii for this circular region, $r = 246$ and $r = 305$ arcsec, to better assess the effect of the background, which was extracted from all the remaining S3 chips. We detected the halo at a $3.8\sigma$ level in the 0.3–7 keV energy range ($1439 \pm 382$ and $884 \pm 239$ counts for the larger and smaller source regions respectively, corresponding in both cases to a surface brightness of $0.011 \pm 0.003$ counts pixel$^{-2}$), and at a $5.5\sigma$ level in the 0.5-2 keV range ($1259 \pm 245$ and $852 \pm 154$ counts for the larger and smaller source regions respectively, corresponding in both cases to a surface brightness of $0.010 \pm 0.002$ counts pixel$^{-2}$). Since the orientation of the chips in the other two observations is different, it is not possible to apply the same regions to them, instead we extrapolated the statistics from the first observation to the total observed time. We thus obtained a $1\sigma$ range for the surface brightness of the halo, in the 0.3–7 keV energy band, of 0.01–0.02 counts pixel$^{-2}$ (0.04–0.08 counts arcsec$^{-2}$).

The constraints on the surface brightness allowed us to test a thermal model with different temperatures, assuming a suitable extraction region outside the radio lobes (a spherical shell centred in the AGN, with $R_{int} = 3.68$ kpc, $R_{ext} = 5.24$ kpc). Given the range of temperatures found in the haloes of similar galaxies (see e.g. Tüllmann et al. 2006b; Yamasaki et al. 2009), we derived the electron densities for $kT = 0.1$ keV, 0.2 keV and 0.3 keV.

We used our low/high values for the surface brightness and the halo shell region to obtain estimated count rates for each temperature. We then derived the *apec* normalizations required to obtain these count rates, and derived the electron densities and pressure from those normalizations. The results are displayed in Table 3. We find that the electron densities we derive are compatible with those estimated by Elmouttie et al. (1998a), based on observations of other nearby spiral galaxies. We also estimated the cooling time for the gas, obtaining results of 0.3–3.6 $\times 10^8$ yr for $kT = 0.2$–0.3 keV, and as low as 6.2 Myr for $kT = 0.1$ keV. These times are very

**Table 2**
Results for the shells of Circinus

| Shell | kT | Norm | $N_e$ | M | P | E | PV | K |
|---|---|---|---|---|---|---|---|---|
| | keV | $\times 10^{-4} \mathrm{cm}^{-5}$ | $\times 10^{-2} \mathrm{cm}^{-3}$ | $\times 10^{36} \mathrm{Kg}$ | $\times 10^{-12} \mathrm{Pa}$ | $\times 10^{54} \mathrm{erg}$ | $\times 10^{54} \mathrm{erg}$ | $\times 10^{53} \mathrm{erg}$ |
| W | $0.74^{+0.07}_{-0.05}$ | $1.39^{+0.24}_{-0.13}$ | $2.27^{+0.11}_{-0.11}$ | $1.97^{+0.09}_{-0.10}$ | $4.98^{+0.14}_{-0.25}$ | $4.56^{+0.22}_{-0.23}$ | $3.04^{+0.14}_{-0.15}$ | $5.15^{+0.25}_{-0.26}$ |
| E | 0.8 | $1.30^{+0.12}_{-0.12}$ | $2.20^{+0.10}_{-0.11}$ | $1.90^{+0.09}_{-0.09}$ | $5.21^{+0.23}_{-0.25}$ | $6.10^{+0.28}_{-0.29}$ | $4.07^{+0.18}_{-0.20}$ | $5.38^{+0.24}_{-0.26}$ |
| E | 1.8 | $1.08^{+0.12}_{-0.12}$ | $2.00^{+0.11}_{-0.11}$ | $1.73^{+0.10}_{-0.11}$ | $10.1^{+0.5}_{-0.6}$ | $5.55^{+0.30}_{-0.30}$ | $8.32^{+0.43}_{-0.45}$ | $4.89^{+0.27}_{-0.26}$ |

**Note.** — These results are for the temperatures discussed in the text and $Z = 0.15\odot$. The columns, from left to right, show the shell, temperature, model normalization, electron density, mass, pressure, total thermal energy, work available from the gas filling the shells and total kinetic energy.

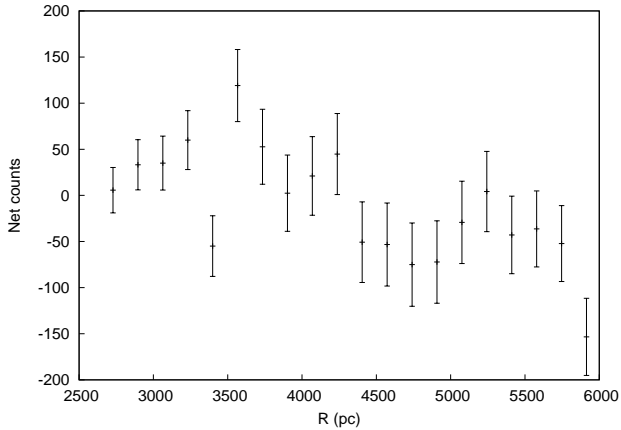

**Figure 8.** Radial profile of background-subtracted counts for the longest *Chandra* observation of Circinus. Scale: 1 pixel = 0.492 arcsec = 9.54 pc. The emission from the halo of Circinus is barely above the background level, but there seems to be a net count drop after $R \sim 4.2$ kpc ($\sim 440$ pixels), indicating that this may be the extent of the halo. Combining the results for the three observations we obtained a surface brightness estimate of $4$–$8 \times 10^{-2}$ counts/arcsec$^2$.

**Table 3**
Results for the external medium.

| kT | Norm | $N_e$ | $L_{0.2-3keV}$ | P |
|---|---|---|---|---|
| keV | $\times 10^{-2}\ \mathrm{cm}^{-5}$ | $\times 10^{-2}\ \mathrm{cm}^{-3}$ | $\times 10^{38}\ \mathrm{erg\ s}^{-1}$ | $\times 10^{-12}\ \mathrm{Pa}$ |
| 0.1 | 2.80/5.70 | 2.35/3.35 | 0.71/1.45 | 0.70/0.99 |
| 0.2 | 1.00/2.10 | 1.40/2.03 | 5.30/10.70 | 0.83/1.20 |
| 0.3 | 0.33/0.66 | 0.80/1.14 | 5.04/10.07 | 0.72/1.01 |

**Note.** — The columns, from left to right, show the low/high values of the model normalization, electron density, luminosity and gas pressure for each assumed gas temperature.

short, since line cooling is very important at these low temperatures, suggesting that the hot gas into which the lobes are expanding must have a comparatively recent origin, possibly in a starburst and/or supernova wind.

To verify our assumptions on the properties of the gas in the halo we used the relations of Tüllmann et al. (2006a), which give the correlation between the X-ray luminosity of the halo and other parameters for a sample of star-forming spiral galaxies. Some of their correlations indicate a very strong dependence between the soft X-ray luminosity of the halo and parameters such as the FIR/H$\alpha$ star formation rate, and between the overall galactic X-ray luminosity and the mass of dust or HI in the galaxy.

Given how obscured and nearby Circinus is, most previous studies have focused just on the central star-forming ring or the supernova remnant, so that some of the parameters we could use to derive the luminosity of the halo are not known, but we were able to use the values from Curran et al. (2008); Jones et al. (1999); Elmouttie et al. (1998b); For et al. (2012) ($L_{B,bol} \sim 2.9 \times 10^{43}$ erg s$^{-1}$, $L_{UV,bol} \sim 9.1 \times 10^{42}$ erg s$^{-1}$, $M_{dust} \sim 10^6\ M_\odot$, $M_{stars} \sim 10^{11}\ M_\odot$, $SFR_{FIR} \sim 3 - 8\ M_\odot$ yr$^{-1}$) with Tüllmann et al.'s correlations. This allowed us to constrain the soft (0.3-2 keV) luminosity of the halo of Circinus to 1.6-6.3$\times 10^{39}$ erg s$^{-1}$ from the correlations of Tüllmann et al. Given that the extraction region we used in our analysis does not cover the full halo of the galaxy, we find that our results are consistent with what is expected from these correlations and the observations (see Table 3). The luminosity we thus constrained is also compatible with the count rates we estimated from our images.

Comparing the electron densities we derive for the shells and the external gas, we see that density ratios of $\sim 1.1$–3 can be derived from $kT_{ext} = 0.2$ or $kT_{ext} = 0.3$ keV, and smaller values (0.6–0.9) for $kT_{ext} = 0.1$ keV. The small density contrast makes the latter temperature less consistent with our shock interpretation, given that the density ratio $\rho_{out}/\rho_{shell}$ tends to 4 in the case of a strong shock. Based on observational results (see e.g. Hummel et al. 1991; Beck et al. 1994), the electron densities we derived from $kT = 0.2$ and $kT = 0.3$ keV are compatible with those seen in the haloes of spiral galaxies.

Although the fact that we do not directly detect the external gas is still a limitation, since it would allow us to rule out a slow shock into a cold, dense medium, or pressure conditions where a shock is not possible, the morphology of the X-ray emission, the observed properties, and the agreement of the extrapolated values for the external gas with what we would expect in our assumed scenario indicate that we are indeed observing a strong shock being driven into the ISM of Circinus.

Using the observed parameters we can calculate the Mach number for the shock, as follows (Landau & Lifshitz):

$$\mathcal{M} = \sqrt{\frac{4(\Gamma + 1)(T_{shell}/T_{out}) + (\Gamma - 1)}{2\Gamma}} \qquad (2)$$

where $\Gamma$ is the polytropic index of the gas, assumed to be 5/3. We find Mach numbers $\mathcal{M} \sim 2.7$–3.6 for the W shell (up to $\mathcal{M} = 5.1$ if the external temperature is closer to $kT = 0.1$ keV) and $\mathcal{M} \sim 2.8$–5.3 for the E shell (up to $\mathcal{M} = 7.6$ for $kT = 0.1$ keV). The sound speed is $\sim 264^{+59}_{-78}$ km s$^{-1}$, and the velocity of the shells is $915^{+45}_{-35}$ km s$^{-1}$ for the W shell and $950^{+470}_{-130}$ km s$^{-1}$ for the E shell. The errors on these values reflect the temperature range of both the shell and the external gas; the uncertainty on the E shell is very large due to the range of possible shell gas temperatures, as indicated in Table

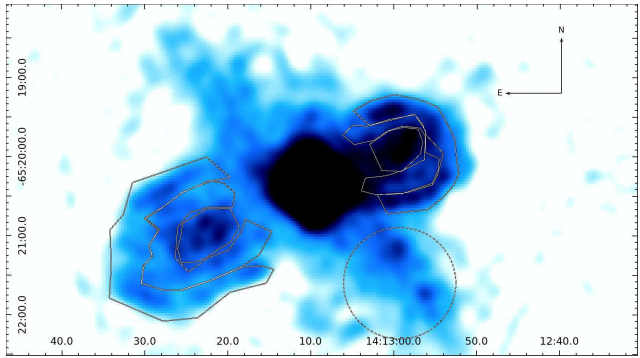

**Figure 9.** 13 cm *ATCA* radio map displaying the regions we used to calculate the spectral index for the different areas (hotspots, lobes and shells, the RMS error was estimated from the dashed circle South of the AGN). With this colour scale the full extent of the radio emission can be appreciated, showing the edge-brightening in the shocked shells, and the dip in luminosity immediately behind them. The resolution of the radio map is 11.8 × 11.0 arcsec.

2. It is clear, in any case, that the velocity of the shells must be supersonic, and close to 1000 km s$^{-1}$, in agreement with what is observed in similar systems (e.g. Mrk 6, NGC 3801: Mingo et al. 2011; Croston et al. 2007)

If we assume that the shells have been expanding with a constant velocity, equal to the one we have derived for the W structure, we can constrain their age to $\sim 10^6$ years. The total energy (enthalpy of the cavity and kinetic energy from the shock) involved in creating both shells is $\sim 2 \times 10^{55}$ erg, the equivalent of $10^4$ supernova explosions with individual energy $10^{51}$ erg, assuming a 100 per cent efficiency.

While the spectral fits of the ionization cone (see Section 4.1) seem to support our initial assumption of $Z = 0.15\odot$, this value is still poorly constrained. It is very likely that the areas of the galaxy covered by our regions contain a multi-phase gas, and a range of abundances, but these cannot be studied in detail without deeper observations. The gas density we derive from our models depends both on the abundance and the emission measure, so that, assuming that there are no abrupt variations in the abundance between the shells and the regions immediately outside them, for $Z = 0.30\odot$ the electron density in the shells would be $\sim 25\%$ lower, while that of the external gas would go down by $\sim 20\%$. The ratio between both densities, however, only depends weakly on $Z$, so even if our assumptions on the abundance were wrong, and the densities were smaller, our conclusions would stand. The Mach number and the shock velocity would not change, but the total energy involved in the process would be slightly smaller.

### 4.3. Starburst-driven bubble

Given that Circinus is a spiral galaxy, with active star formation, it is possible that the bubbles that we are observing perpendicular to the plane of the galaxy may be caused by a starburst or supernova-driven wind. As pointed out by Elmouttie et al. (1998a), however, the particle component within starburst-driven "bubbles" is expected to be a mixture of thermal and non-thermal electrons, with the thermal material causing depolarization. The W "bubble" in Circinus does not show a high degree of depolarization. In the E "bubble" the higher depolarization is most likely caused by the gas in the disk, which is obscuring part of the structure (for full details on the radio polarization of Circinus see Elmouttie et al. 1998a).

As we saw in Section 4.2, the total energy involved in the creation of both "bubbles" in Circinus is $\sim 2 \times 10^{55}$ erg. This value is similar to the one Matsushita et al. (2005) calculated for the superbubble observed in M82 (0.5–2×10$^{55}$ erg), one of the most powerful examples of superbubbles. The efficiency with which supernova explosions transfer energy to the ISM is quite low (3–20 per cent, depending on the density of the ISM and time lapse between subsequent explosions). Elmouttie et al. (1998a) inferred a supernova rate of 0.02 sn yr$^{-1}$ from their radio observations. To check this value we used the far-infrared data from Genzel et al. (1998) ($L_{FIR} = 7 \times 10^9 L_\odot$) and the correlations from Condon (1992) to estimate a massive star formation rate ($SFR_{M \geq 5M_\odot} \sim 0.64 \ M_\odot \ yr^{-1}$) and derived a supernova rate of $\sim 0.026$ sn yr$^{-1}$. With either of these values, the possibility of a supernova-driven wind is energetically viable in Circinus. However, the enhanced emission at the centre of the "bubbles" is not easy to explain, morphologically, with this model. Multiple generations of "bubbles", within the $10^6$ yr age estimated for the larger structure, seem unlikely.

Some consideration must be given to how the radio "bubbles" are related to the structures closer to the AGN. Elmouttie et al. (1998b) suggest that there is an ongoing outflow of gas in the W ionization cone (with a total kinetic energy of 5×10$^{53}$ erg), which could be powered by the AGN or a starburst wind. They suggest that the similar sizes of the $H\alpha$ central ring and the base of the central radio emission indicate that these structures are related, and that an ongoing starburst wind is the cause of both the inner outflow and the enhanced radio emission at the centre of the "bubbles". Emission line ratios in the cone, however, suggest that the emission in the cone may be caused by AGN-driven photoionization. For Circinus, Elmouttie et al. also point out that the morphologies of the ionization cone and the radio "plume" are quite different, indicating that these structures may not have the same origin. As suggested by Schulz (1988) for NGC4151, it is possible that, as the shock that created the larger structures in Circinus made its way through the halo gas, it created a low-density region through which ionizing radiation, from the AGN or a starburst-wind, could escape, giving rise to the cone. It is thus entirely possible that there is an ongoing starburst wind in the central regions of Circinus, feeding material into the ionization cone, but that this is not powerful enough to be the cause behind the large-scale emission in the shells and at the centre of the "bubbles".

The strongest argument against the starburst-driven scenario as an explanation for the large-scale emission, however, comes from the variations in the radio spectral index we observe across the "bubbles". Variations in the spectral index are a sign of a change in the electron population; a lower value of $\alpha$ may indicate a region where particle acceleration is occurring. An increasing value of $\alpha$ indicates radiative losses and an aging electron population. We calculated the fluxes for the shells, the areas immediately behind the "bubbles", and the regions with enhanced radio and X-ray emission inside them (which we labelled as hotspots), as illustrated in Figure 9. We convolved the 6 cm, 13 cm and 20 cm *ATCA* radio maps so that the spatial resolution was the same in all of them (20.5 × 20.5 arcsec), thus ensuring that we are measuring the emission from the same structures in all cases. To best estimate the extent of the RMS noise we chose a background region over the galaxy's disk, where there are flux variations, rather than a region off-source. We then derived spectral index values for each structure from the 6 cm/13 cm and from the 13 cm/20 cm flux measurements. The results are illustrated in

| Structure | $\lambda$ | $S$ | $\alpha$ |
|---|---|---|---|
| | cm | mJy | |
| W hotspot | 20 | $29.0 \pm 2.0$ | $0.62 \pm 0.09$ |
| | 13 | $20.7 \pm 1.4$ | $0.59 \pm 0.05$ |
| | 6 | $13.6 \pm 0.7$ | |
| E hotspot | 20 | $20.9 \pm 3.3$ | $0.60 \pm 0.13$ |
| | 13 | $15.1 \pm 1.7$ | $0.96 \pm 0.07$ |
| | 6 | $7.7 \pm 0.8$ | |
| W shell | 20 | $36.9 \pm 3.3$ | $0.70 \pm 0.13$ |
| | 13 | $25.3 \pm 2.3$ | $0.94 \pm 0.08$ |
| | 6 | $13.0 \pm 1.0$ | |
| E shell | 20 | $37.6 \pm 4.8$ | $0.84 \pm 0.20$ |
| | 13 | $23.9 \pm 3.4$ | $1.12 \pm 0.12$ |
| | 6 | $10.8 \pm 1.5$ | |
| W lobe | 20 | $35.3 \pm 2.6$ | $0.71 \pm 0.11$ |
| | 13 | $24.1 \pm 1.8$ | $0.77 \pm 0.06$ |
| | 6 | $14.0 \pm 0.8$ | |
| E lobe | 20 | $37.2 \pm 3.9$ | $0.78 \pm 0.16$ |
| | 13 | $24.3 \pm 2.7$ | $0.97 \pm 0.09$ |
| | 6 | $12.2 \pm 1.2$ | |

**Note**. — The spectral indices for the different extended structures (hotspots, shells and lobes) of Circinus, have been derived from the *ATCA* 6 cm, 13 cm and 20 cm maps. The regions we used to calculate the relative fluxes are illustrated in Fig. 9.

Table 4.

Although our results are limited by the very low flux levels and by the fact that at 6 cm the structures we are observing are slightly different, it is clear that the spectral index does vary between these three regions, and consistently so for both the Western and Eastern structures, in the sense that there is flattening (*alpha* decreases) in both hotspots and steepening (*alpha* increases) in both shells, even if the individual values are slightly different (this may be attributed to low spatial resolution, disk obscuration of the E lobe, and/or different spectral indices in the pre-shock electrons for both shells). These results are also consistent with the spectral index map from Elmouttie et al. (1998a). As stated above, this symmetrical flattening of the spectral index at the centre of the "bubbles" is most easily explained if there is ongoing particle acceleration. The simplest scenario to explain this is that these regions are points where a radio jet interacts with the ISM, and thus the equivalents of the hotspots observed in more powerful systems (see Section 4.4 for more details). It is natural to assume that this same jet is the one that provided the energy injection that caused the shock we are observing. This is further supported by the morphology of the emission, which is coincident in X-rays and radio, and symmetrical for both lobes, roughly equidistant from the nucleus of the galaxy.

### 4.4. *Jet-driven shocks*

In light of these results we find that the scenario that best describes the data is that which is typically assumed for radio galaxies (see e.g. Scheuer 1974; Kaiser & Alexander 1997). In this scenario the jet emanating from the central AGN makes its way into the ISM of the host galaxy, transferring part of its energy to the surrounding gas at the termination points (or hotspots). This energy transfer heats up the ISM gas, creating a supersonic bow shock. The material that has flowed up the jet is observed in the form of radio lobes.

Hotspots in X-rays are characterised by a non-thermal powerlaw spectrum, though they are generally very faint, given that synchrotron losses are very high at these energies. As we stated in Section 3.2, the statistics of the spectra of the hotspot regions are quite poor, and they do not allow us to distinguish between thermal and non-thermal emission. If the emission were non-thermal, it would be unlikely to have its origin in widespread inverse Compton scattering, since the contribution of such a component for low-power radio galaxies is negligible. We estimate the IC luminosity density at 1 keV to be $\sim 6 \times 10^{34}$ erg s$^{-1}$ keV$^{-1}$ for the case of Circinus (equivalent to $2$–$5 \times 10^{-5}$ counts s$^{-1}$ in *Chandra* ACIS-S, depending on the IC powerlaw photon index), using the assumptions on the $B$ field detailed in Section 4.5. Moreover, if there was a detectable contribution from inverse Compton it would appear uniformly across both lobes. We calculated the spectral index between 6 cm and 1 keV for the W hotspot, and found it to be $\sim 1$. This value is in the range of radio/X-ray ratios seen in FR I jets (see e.g. Harwood & Hardcastle 2012), suggesting that synchrotron emission could account for at least part of the X-ray emission we are observing.

Although our images do not show an active jet (at either radio, optical or X-ray frequencies), we know that this does not rule out the possibility that there is one, from what is observed in powerful FR II radio galaxies, where the jets are rarely seen, yet the presence of hotspots clearly indicates their presence (e.g. Kharb et al. 2008). The jet's rest-frame emissivity may be very low in Circinus. Moreover, the orientation of the lobes suggests that the jet should be at a large angle to the line of sight, so that relativistic beaming may be reducing the visibility of a jet that would intrinsically be visible. Doppler-dimmed jets and hotspots have been found in other Seyfert galaxies with low radio luminosities (see e.g. Kukula et al. 1999), illustrating how the interactions of the jet with different, complex environments can change the observed radio properties of these sources. This might also explain why the X-ray emission in the shells extends further than the radio along the (presumed) direction of propagation of the radio jet. This geometry is expected from some jet propagation simulations (see e.g. Sutherland & Bicknell 2007).

In this context, to further verify our shock scenario, we calculated the minimum internal pressure of the radio lobes under equipartition conditions from the radio data, fitting a broken power-law electron energy spectrum with $p = 2$ at low energies, steepening to $p = 3$ at the electron energy that gives the best fit to the data. We assumed ellipsoidal geometry, no protons and $\gamma_{min} = 10$. We obtained a value of $P \sim 6$–$9 \times 10^{-14}$ Pa, consistent with the results obtained by Elmouttie et al. (1998a), and roughly 2 orders of magnitude lower than the pressures we derive from our X-ray data. This departure from minimum energy is often found in FR I radio galaxies (e.g. Morganti et al. 1988); we also found this effect in NGC 3801, Mrk 6 and Cen A. These results imply that there is some additional contribution to the internal pressure, caused by a large deviation from equipartition conditions, a high fraction of non-radiating particles, such as thermal or relativistic protons originating from the interaction of the galaxy's gas with the jet, or a low filling factor. Of these, the second explanation is the most plausible one (see Hardcastle et al. 2007; Croston

et al. 2008a).

From the age of the shells and the total energy we can infer the kinetic luminosity of the (invisible) jet, which is $\sim 10^{41}$ erg s$^{-1}$. Comparing these results to those we obtained for Mrk 6 (Mingo et al. 2011), NGC 3801 (Croston et al. 2007) or Cen A (Kraft et al. 2003) shows that these parameters scale down with the radio power of the parent AGN (see Table 5).

The fact that the X-ray emission in the shells is thermal is expected, if our shock scenario is correct. As pointed out by Croston et al. (2009), it is likely that fast shocks are required to produce in-situ particle acceleration and X-ray synchrotron emission in the shocked shells, as observed in Cen A. The velocities we derive, smaller than those found in Cen A, and the unequivocally thermal nature of the emission we observe, suggest that the case of Circinus is more similar to that of NGC 3801 or Mrk 6.

### 4.5. Implications for particle acceleration and parallels with SNR

The fact that the radio and X-ray emission are spatially coincident in the shells of Circinus resembles the morphology that is observed in some SNR, and poses the question of how similar the underlying physical processes are. Given that we do not observe non-thermal emission in the X-rays in the shells of Circinus, but we do see edge-brightening in radio, we investigate whether this enhancement of the radio synchrotron emission in Circinus can be fully accounted for through a shock-driven enhancement of the $B$ field strength, similar to what has been observed in ongoing cluster mergers (e.g. Markevitch et al. 2005; van Weeren et al. 2010). The configuration of the magnetic field in Circinus is not very clear. The argument of Elmouttie et al. (1998a), stating that the dominant component of the magnetic field in the shells of Circinus seems to be predominantly radial (perpendicular to the shock front, parallel to its direction of propagation) is based on polarization maps and is not very strong. Other systems such as NGC 1068 (Wilson & Ulvestad 1987; Simpson et al. 2002) and NGC 3079 (Duric & Seaquist 1988) indicate that fields tangential to the shock front (perpendicular to the direction of propagation of the shock) are more often found in radio galaxies. Both radial and tangential $B$ field configurations can be found on SNR (see e.g. Milne 1987), but only the component that is tangential to the shock front is compressed by it. Compressed magnetic fields have been found around the leading edges of FR I radio galaxies (Guidetti et al. 2011, 2012) from rotation measure analysis, suggesting that, given the right conditions (a component to the magnetic field tangential to the front and a strong shock) this phenomenon may be common to many low-power systems.

To investigate whether this shock-driven $B$ enhancement could be enough to account for the enhancement in radio emission, we must calculate whether the jump in emissivity (of a factor of at least 40) observed between the shells and the gas just outside them can be explained by this mechanism. The synchrotron emissivity of a powerlaw distribution of electrons is given by :

$$J(\nu) = C N_0 \nu^{-\frac{(p-1)}{2}} B^{\frac{p+1}{2}} \qquad (3)$$

where $C$ can be assumed constant, $N_0$ is the normalization of the relativistic electron spectrum, and $(p-1)/2 = \alpha$. If we assume that $B$ is tangential to the shock front, applying the Rankine-Hugoniot conditions for a strong shock results in $B_{shells}/B_{ext} \propto \rho_{shells}/\rho_{ext}$, which, substituting the values from Tables 2 and 3, results in $B_{shells} \sim 1.1-3 B_{ext}$. In Eq. 3 $N_0$ is proportional to the electron density, but given that the electron density depends on the energy density, which scales with the temperature in the shock, we have that $N_{0,shells}/N_{0,ext} \propto (\rho_{shells}/\rho_{ext})(T_{shells}/T_{ext})^{p-1}$. Substituting the values from Tables 2, 3 and 4 we can calculate the emissivity jump. Taking only the values for the W shell, which are better constrained, we see that $\alpha \sim 0.70-0.94$ and $T_{shell} \sim 0.69-0.81$ keV. If $T_{ext} = 0.2$ keV, using the densities from Tables 2 and 3, we have ($\rho_{shells}/\rho_{ext}$) $\sim 1.1-1.7$, and ($T_{shells}/T_{ext}$) $\sim 3.5-4.1$, thus ($J_{shells}/J_{ext}$) $\sim 7-68$. If $T_{ext} = 0.3$ keV we have ($\rho_{shells}/\rho_{ext}$) $\sim 1.9-3.0$, and ($T_{shells}/T_{ext}$) $\sim 2.3-2.7$, thus ($J_{shells}/J_{ext}$) $\sim 18-164$.

These results show that shock-compression alone could be enough to fully account for the enhancement in radio synchrotron emission if the conditions are favourable (a strong shock and a magnetic field tangential to the shock front). It is likely that, even in the best conditions, there is a component of the magnetic field that is perpendicular to the shock front. Our results do not preclude local particle acceleration—it is in fact needed if the conditions at the shock are not ideal—but we know that any particle reacceleration is not efficient enough to power synchrotron emission in the X-rays.

When making the parallel between the shells of Circinus and what is observed in SNR, some caution must be exercised. In young SNR where $B$ field amplification is observed, it is mediated by efficient particle acceleration (Völk et al. 2005). It seems to be clear that, in most cases, field compression alone cannot account for the acceleration of cosmic rays that is observed in SNR, which are in fact the main source of Galactic cosmic rays. The field amplification in SNR is much larger than the enhancement expected from shock compression alone, given that the values of $B$ in SNR tend to be of the order of mG (Vink & Laming 2003), 2–3 orders of magnitude higher than the values found in spiral galaxies, where typical values for the ISM of the disk are $\sim 10$ μG, (see e.g. Beck et al. 1996; Vallée 2011; Hummel et al. 1991; Beck et al. 1994). Synchrotron losses also become substantial at the values of $B$ found in most SNR. Rayleigh-Taylor instabilities at the shock front (Guo et al. 2012) seem to be present in most SNR. While Rayleigh-Taylor instabilities may be behind some of the $B$ enhancement in shock shells around radio galaxy lobes, strong instabilities are not expected to develop until the jet has switched off and the radio lobes have become buoyant (see e.g. Brüggen & Kaiser 2001).

The fact that the spectral index in the hotspots of Circinus is flatter than that of the lobes suggests that there is spectral ageing of the synchrotron emission, evolving radially outwards from the hotspots. In the model we discuss here the electron population in the shells, however, must come from electrons from the external medium swept up in the shock. These electrons are already relativistic, as we know from synchrotron observations of the haloes of spiral galaxies (see e.g. Beck et al. 1996). We cannot directly measure the spectral index of the emission outside the shells, since it is too faint. However, we studied the spectral index in the disk, and found a value $\alpha \sim 1$. This value is similar to that found in the shells, hinting that this is the spectral index of the electron population outside the shells as well. This provides additional support to the idea that the halo electrons swept up in the shock are not strongly reaccelerated, since that would flatten their spectral index, and that the main mechanism behind the enhanced synchrotron emission in the shells we observe in the radio maps is indeed a $B$ field enhancement. The thermal nature of the X-ray emission in the shells also supports the hypothesis of shock compression as the dominant mechanism behind the radio emission enhancement.

We carried out equipartition calculations for the shells, and found that, under those conditions, the pressures we derive are over an order of magnitude below our measurements. This suggests that the shells are not in equipartition and that the contribution to the total energy density from the synchrotron-emitting electrons is not the dominant one. This is to be expected, since the spatial coincidence between the X-ray emitting hot gas and the synchrotron-emitting plasma implies that they coexist in this region, and the former component is expected to dominate the energy density.

As mentioned above, compressed magnetic fields may be common to low-power radio galaxies. Given that field enhancement through shock compression does not preempt particle acceleration, it is likely that in most systems a combination of both is responsible for the observed emission. An accurate measurement of the spectral index and emissivity of the electrons outside the lobes, and better X-ray statistics for both the shells and the external environment, would allow us to estimate the contributions from both processes in Circinus and in other systems. In the case of systems more similar to Cen A, where particle acceleration seems to be causing synchrotron emission in the shells, all the way to the X-rays, constraints on the magnetic field in the shock, e.g. via rotation measure analysis, as in Guidetti et al. (2011), could be used for estimations on the contribution to the emission caused by magnetic field compression.

Our results suggest that the parallel between the jet-driven shocks of low power FR I galaxies and those of SNR, where a similar morphology and some of the same physical processes can be expected, is more common than we previously thought. The fact that it has only been observed in Circinus is mainly due to the small distance between this galaxy and our own, which allows us to spatially resolve these structures, and observe their characteristics with the level of sensitivity of our current instruments. This is particularly relevant for more distant low-power radio galaxies in general, and those with late-type hosts in particular. It is also significant for the edge-brightening in the radio lobes of Mrk 6 (Kharb et al. 2006; Mingo et al. 2011), which could be explained through this mechanism. As pointed out in Section 4.2, the shock results suggest that we are facing a scenario similar to the one we found in NGC 3801, suggesting that some of our conclusions may apply to this object, as well as, possibly, to NGC 6764.

## 5. CONCLUSIONS

We have presented a detailed analysis of the X-ray emission associated with the radio lobes of the Circinus galaxy.

Our deep *Chandra* observation has allowed us to study both the morphological correspondence between radio and X-ray emission, and the X-ray spectral properties of the gas in these structures. We conclude that we are observing shells of shocked gas around the radio lobes of Circinus, expanding into the halo regions of the galaxy with Mach numbers $\mathcal{M} \sim 2.7$–3.6 for the W shell and $\mathcal{M} \sim 2.8$–5.3 for the E shell, consistent with the Rankine-Hugoniot conditions for a strong shock.

We rule out the possibility that this emission is caused by AGN photoionization, and discard the scenario in which the radio structures are the result of star formation or a supernova-driven wind; instead we argue that they are caused by the interaction of a radio jet with the surrounding gas. The total energy (thermal and kinetic) involved in the creation of these shells is $\sim 2 \times 10^{55}$ erg. We estimate their age to be $\sim 10^6$ years. From these parameters we infer the kinetic luminosity of the jet, which is $\sim 10^{41}$ erg s$^{-1}$. When comparing these results with those previously obtained for Cen A, Mrk 6 and NGC 3801, we observe that they scale with the radio power of the source.

The presence of two symmetrical structures resembling jet termination points or "hotspots", together with the slightly different X-ray spectra and flattening of radio spectral index in these regions, suggests that the radio jet may still be active, even if it is not visible. This suggests that "invisible" jets may be a more common occurrence among Seyfert and late type galaxies than previously thought, given that it is difficult to separate the different radio structures in galaxies with active star formation, when the radio bubbles are faint or their orientation is similar to that of the disk. This opens new questions on the role of low radio luminosity, star-forming, active galaxies, which are often ignored in survey studies.

In Circinus the X-ray and radio emission from the shocked shells are spatially coincident. This, added to the fact that the radio emission is edge-brightened, allows us to draw a parallel between these structures and what is observed in supernova remnants: the radio emission from the shells comes from shock-compressed or (re)accelerated cosmic ray electrons in the environment, but particle acceleration, if present at all, is not efficient enough to produce TeV electrons, so that only thermal emission is observed in X-rays. This correspondence has been long expected for AGN, given the similarity between the physical processes involved in SN and jet-driven shocks, but has not been unequivocally observed before. We believe this scenario may be common to other low-power radio galaxies, and potentially very relevant to those with late-type hosts, but can only be observed where we are able to spatially resolve the different structures, something that has been achieved for Circinus due to its proximity to us.

### ACKNOWLEDGEMENTS

BM thanks the University of Hertfordshire for a PhD studentship. JHC acknowledges support from the South-East Physics Network (SEPNet). This work has made use of new data from *Chandra* and software provided by the Chandra X-ray Center (CXC) in the application package CIAO. The authors wish to thank Marc Elmouttie for providing his original radio maps, Bi-Qing For for providing the $H\alpha$ maps and preliminary *Spitzer* results on the SFR, as well as the students from the ATNF Summer school (2009) for making available *ATCA* data obtained through their summer project. This paper includes archived data obtained through the Australia Telescope Online Archive (http://atoa.atnf.csiro.au). We thank the anonymous referee for the useful comments.